# SEGMENTATION OF MULTIPLE SCLEROSIS LESION IN BRAIN MR IMAGES USING FUZZY C-MEANS

Saba Heidari Gheshlaghi[1], Abolfazl Madani[2], AmirAbolfazl Suratgar[3] and Fardin Faraji[4]

[1]Department of Electrical Engineering, Amirkabir University of Technology (Tehran Polytechnic), Tehran, Iran
[2]Department of Control Engineering, South Tehran Branch Islamic Azad University (IAU) Tehran, Iran

[3]Department of Electrical Engineering, Amirkabir University of Technology (Tehran Polytechnic), Tehran, Iran
[4]Neurology Department, Arak University of Medical Sciences, Arak, Iran

*ABSTRACT*

*Magnetic resonance images (MRI) play an important role in supporting and substituting clinical information in the diagnosis of multiple sclerosis (MS) disease by presenting lesion in brain MR images. In this paper, an algorithm for MS lesion segmentation from Brain MR Images has been presented. We revisit the modification of properties of fuzzy-c means algorithms and the canny edge detection. By changing and reformed fuzzy c-means clustering algorithms, and applying canny contraction principle, a relationship between MS lesions and edge detection is established. For the special case of FCM, we derive a sufficient condition and clustering parameters, allowing identification of them as (local) minima of the objective function.*

## 1. INTRODUCTION

The importance of finding a correct and efficient way for diagnosing diseases are increased over time and it is important to find an optimal and accurate method for diagnosing diseases. Multiple sclerosis, also known as MS, is a chronic disease that attacks the central nervous system (CNS) and affects white matter by the person's own immune system so that MS is known as an auto-immune disease. A nerve fiber is surrounded by myelin, which protects the nerves and helps them to conduct electrical Signals (Impulse) between each other. Myelin is vital to the normal functioning of the nervous system. When a nerve carries electrical impulses from one end to the other, myelin helps to stop the impulse from leaving the axon and increases electrical resistance, therefore enhancing signal conduction. When the myelin is disappeared which is called Demyelination, a variety of cognitive, motor and sensory deficits appears and can cause neurological disease. One of the most well-known demyelinating diseases is MS. MS is a largely unknown disease these days and diagnosis it correctly and early, has a significant impact on disease progression [1].

The demyelination makes lesions in the brain. Hence, Brain lesions detection plays an important role in MS studies, as it is used to evaluate patient disease and, it is a future development. Currently, lesions are detected manually or semi-automatic segmentation methods, which are very time-consuming and show a high inter and variability [2].

DOI : 10.5121/ijaia.2018.9203          37

Magnetic Resonance Imaging (MRI) was officially included in the diagnostic of patients presenting with a clinically isolated syndrome suggestive of multiple sclerosis in 2001 by an international panel of experts [3]. Diagnosing multiple sclerosis depends on the sign of disease dissemination in space and time and exclusion of other syndromes that can imitate multiple sclerosis by their clinical and laboratory Specifications. MRI can support and substitute clinical information for multiple sclerosis diagnoses, allowing an early and accurate diagnosis. [4] MR images is a valuable application and most common method for identification MS patients. The clinical presentation of MS includes a wide range of physical disorders and cognitive symptoms. Cognitive Impairment decrease quality of life and treatment amongst MS patients [5-8]. MS is an inflammatory demyelinating and degenerative disease of the CNS, distinctive pathologically by a different part of brain inflammation, demyelination, axonal loss, often causing motor, sensorial, vision, correlation, and cognitive impairment [9].

Although, MS is the recurring, neurological disease proficient in disability in young adults. In these days, MS disease has been increased, and geographical areas play a significant role in MS. Moreover, Multiple Sclerosis is between two and three-times more average in females than in males, but males have a propensity for later disease [10]. As mentioned, one of the most famous facilities for diagnosis MS disease is Magnetic Resonance Imaging (MRI) techniques [9], T2-weighted (T2-w) and gadolinium-enhanced T1-weighted (T1-w), are highly aware in detecting MS lesions. MRI-derived metrics have become the most important medical tool for diagnosing MS disease.

Both sharp and persistent MS plaques appear as focal high-signal intensity areas on T2-w sequences, all tissues illustrated by water content. The total T2 lesion volume of the brain increases by almost (5–10%) each year in the relapsing forms of MS [11]. Gadolinium-enhanced T1-w imaging is highly fragile in detecting inflammatory activity. CNS atrophy, which involves both grey matter and white matter, is a progressive phenomenon that becomes worse with argument disease length and progresses at a rate of between 0.6% and 1.2% of brain loss per year in this disease [12]. It is important to consider that different segmentation methods might have an impression on the small volume of the lesion detection and this small volume of the lesion can be used to show the disease progression.

In this paper, Section 2 reviews of the pre-processing steps needed automatically portioning MS lesions. Section 3 approaches segmentation methods and also reviews algorithms which are used. In Section 4, conclusion and results are showing and give some ideas for future works.
In this research, MRIs for patients with MS are from various sources. Mostly, our high gratitude should be directed to C.P. Loizoua and his colleagues: The Laboratory of Health at the University of Cyprus (http://www.medinfo.cs.ucy.ac.cy/)

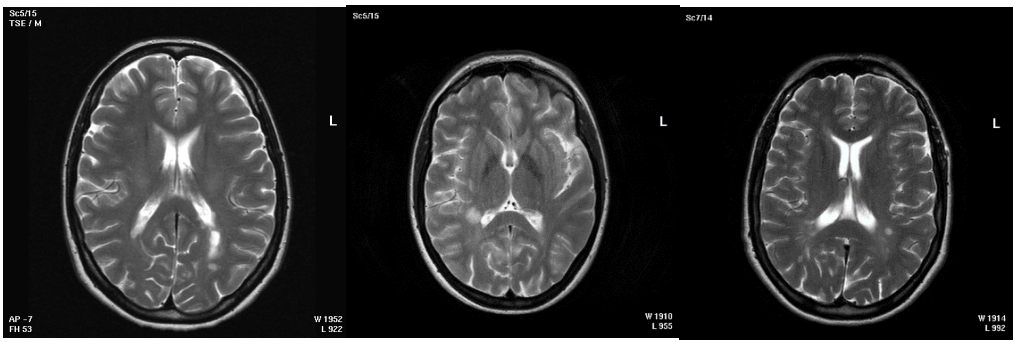

Figure 1: Samples Of Mr Images For Ms Patient

## 2. PRE-PROCESSING STEPS

Accurate and efficient association of brain in MR head images is a critical first step in near all neuroimaging studies. There are different methods according to skull extraction . Although, this section is significant, using an efficient method. In this research, the automated brain extraction tool (BET) (Smith, 2002) has been used [13-16]. BET is an automated Brain Extraction Tool designed to segment the brain and non-brain tissues in MRI head scan volumes as well, image binarization is measuring by Eq. (1). th is described as a threshold.

$$Bin(x,y) = \begin{cases} 1 & if \quad f(x,y) > th \\ 0 & if \quad Ow \end{cases} \quad (1)$$

## 3. SEGMENTATION METHODS AND ALGORITHM

For automatic analysis and segmentation images, varieties of methods have been generated during last three decays and so images could be segmented into its basic elements. Many edge detection methods are used to find edges of objects in the images. Edges mostly occur on the boundary between two different regions in the image.

Edge detection is an image processing technique for finding the boundaries of objects by detecting brightness discontinuities. Image edge detection methods have been studied intensively for the last three decades, and it has an important role in image processing and computer vision systems. In addition, it is used concerning image segmentation, data extraction like image processing, computer vision, and machine vision, etc.
Various classification methods for edge detection has been cleared, but they are usually divided into two main groups:

- Gradient based method: The gradient method looks for maximum and minimum in the first derivative of the image for detecting the edges [17].
- Laplacian based method: It searches for zero crossings in the second derivative of the image for finding edges. An edge has the one-dimensional shape of a ramp and calculating the derivative of the image can highlight its location. The zero crossing of the distinguished signal as edge points [17]. Because images are two-dimensions, the two-dimensional Gaussian operator has been applied. The two-dimensional Gaussian operator G(x, y) is defined:

$$G(x,y) = \frac{1}{2\pi\sigma_x\sigma_y} e^{-(\frac{x^2}{2\sigma_x^2}+\frac{y^2}{2\sigma_y^2})} \quad (2)$$

If we considering these condition; i.e. $\sigma_x = \sigma_y$ and $x^2 + y^2 = r^2$ the equation simplify to:

$$G(r) = \frac{1}{2\pi\sigma^2} e^{\frac{r^2}{2\sigma^2}} \quad (3)$$

For edge detection, searching for zero-crossing in 2nd order derivative of the image needed which f(x,y) is the image [2].

$$g(x,y) = \nabla^2 G(r) * f(x,y) \tag{4}$$

$$\nabla^2 G(r) = \frac{-1}{4\pi\sigma^2}[1 - \frac{r^2}{2\sigma^2}]e^{\frac{r^2}{2\sigma^2}} \tag{5}$$

The most common edge detection algorithms are Sobel, Canny, Prewitt, Marr-Hildreth. Those methods are determined in this research. The main difference between these methods is that canny edge detection method is in the spatial domain while others are in the frequency domain [9].

In Figure 2, the differences between Canny, Sobel, and Marr-Hildreth edge detection is shown. As it shown in figure 2, canny edge detection has a better result than other methods. The Canny edge detector is one of the best edge detections method, which is currently used, also canny edge detector has good noise reduction and at the same time, it detects true edge with the minimum error [18] [19].

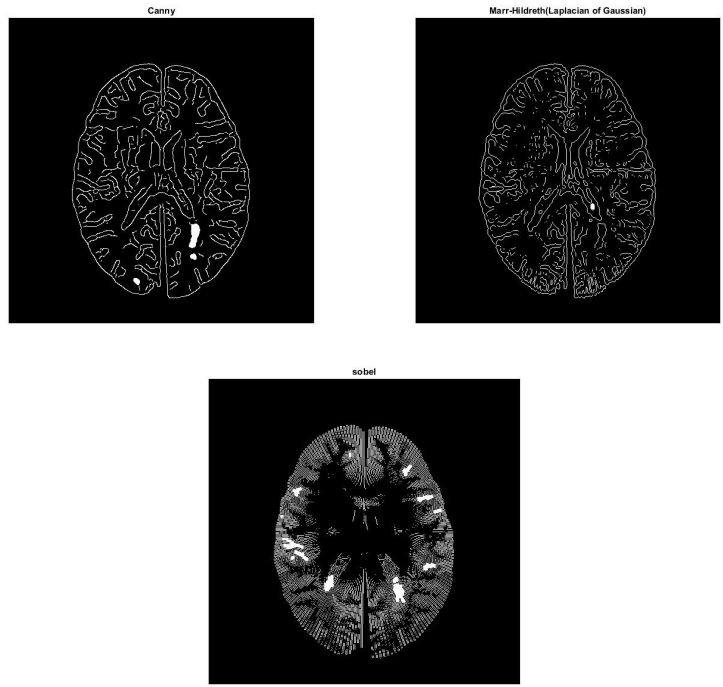

Figure 2: Differences between Canny, Sobel, and Marr-Hildreth edge detection methods.

## 4. FUZZY C-MEANS

Clustering of numerical data forms of many classification and system modeling methods. The suggestion of clustering is to identify natural groupings of data from a large dataset to create a concise representation of a system's behavior. Semi-supervised classification algorithms based on fuzzy c-means (FCM) clustering in which a data set is grouped into n clusters with any data point in the data set belonging to every cluster to the main degrees [20]. The FCM algorithm assumes that each spike concerns to more than one cluster with given a dataset $X = \{x_1, x_2, x_3..., x_N\}$ clustering goals to discrete the dataset into alternating subsets each with a cluster midpoint. FCM algorithm allocates fuzzy memberships $u_{ij}$ any pixels $x_j$ ($j = 1, 2, ..., N$) in each variety c by decreasing the following cost function:

$$J = \sum_{j=1}^{N} \sum_{i=1}^{C} u_{ij}^{m} \| x_j - v_i \|^2 \tag{6}$$

Where m is the weighting fuzziness parameter. $v_i$ Illustrate the i-th cluster midpoint. $\|.\|$ Explain the Euclidean interval. $u_{ij}$ Means a partition matrix that is subject $u_{ij} \in [0,1]$ $and$ $\sum_{i=1}^{c} u_{ij} = 1$. Using the Lagrangian method, partition matrix, and cluster centers by calculating as follows:

$$v_i = \frac{\sum_{j=1}^{N} u_{ij}^{m} x_j}{\sum_{j=1}^{N} u_{ij}^{m}} \tag{7}$$

$$u_{ij} = \frac{1}{\sum_{k=1}^{c} (\frac{\| x_j - v_i \|^2}{\| x_j - v_k \|^2})^{\frac{1}{(m-1)}}} \tag{8}$$

In this study, considering the MR image is often with spatial inhomogeneity due to the deficiency in the magnetic field, the intensity $x_j$ is modeled by $x_j = y_i - \gamma_j$ here, $y_i$ and $\gamma_j$, and mean the measured absorption and the consistent bias field. We propose a novel algorithm as an addition of the traditional FCM clustering by modifying the cost function in Equation 9 as [21, 22]:

$$J = \sum_{j=1}^{N} \sum_{i=1}^{c} u_{ij}^{m} \| x_j - v_i \|^2 + \alpha S + \beta R \tag{9}$$

Where S is a normalizer that indicate the neighborhood influence during segmentation. R is the regularization term on bias field. $\alpha$ and $\beta$ are constants that control the effect of the two regularization term correspondingly definitely, S, R are established as [14]:

$$S = \sum_{j=1}^{N} \sum_{i=1}^{C} u_{ij}^{m} (\frac{1}{N_\varepsilon} \sum_{x \in N_\varepsilon} \| x - v_i \|^2)$$

and

$$R = \sum_{j=1}^{N} \sum_{i=1}^{C} u_{ij}^{m} (\| \gamma_j \|^2) \tag{10}$$

Where $N_\varepsilon$ represent neighborhood positioning at $x_j$.

Also in the traditional FCM algorithm, by differentiating the cost function respect to, $V_i$ and $\gamma_j$ respectively and setting the result to zero, we can obtain the updating $u_{ij}^*$, $v_i^*$ and $\gamma_j^*$ as:

$$u_{ij}^* = \frac{1}{\sum_{k=1}^{c}\left(\frac{\|x_j - v_i\|^2 + \alpha.\frac{1}{N_\varepsilon}\sum_{x \in N_\varepsilon}\|x - v_i\|^2 + \beta.\gamma_j}{\|x_j - v_k\|^2 + \alpha.\frac{1}{N_\varepsilon}\sum_{x \in N_\varepsilon}\|x - v_k\|^2 + \beta.\gamma_j}\right)^{\frac{1}{(m-1)}}} \quad (11)$$

$$v_i^* = \frac{\sum_{j=1}^{N} u_{ij}^m \left(x_j + \alpha.\frac{1}{N_\varepsilon}\sum_{x \in N_\varepsilon}(x - v_i)\right)}{(1+\alpha)\sum_{j=1}^{N} u_{ij}^m} \quad (12)$$

$$\gamma_j^* = y_j - \frac{\sum_{i=1}^{c} u_{ij}^m v_i}{(1+\beta)\sum_{i=1}^{c} u_{ij}^m} \quad (13)$$

The final result is shown in Figure 3. For having this result, earliest, we use the pre-processing method as described in section 2, after that applying canny edge detection and finally, for increasing accuracy, Fuzzy C-means has been applied.

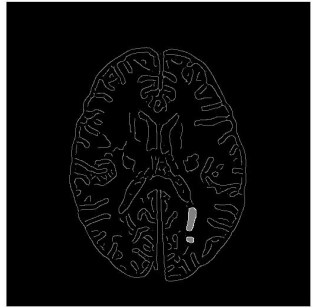

Figure 3: Result when FCM applied after the canny edge detection.

## 5. CONCLUSIONS

Multiple sclerosis caused when the immune system attacks myelin, which covered around human nerve fibers to protect them and help them to send the messages between brain and body efficiently [1]. Without this shell, nerves became damaged. This damage is called Demyelination. Demyelination can make lesions which form MS, and these lesions make the MS a long-lasting disease that can affect the brain, spinal cord, and the optic nerves in eyes. By this damage, transferring messages between brain and body will be affected and symptoms such as fatigue, numbness, vision problem, cognitive problems, etc. appear [23]. The message will slow down or blocked during transferring and leading to the symptoms of MS. No two cases of MS are exactly same, and symptoms are different from one patient to other and it depends on where myelin plaques have formed. Because of these differences, Doctors use medical history, physical exam, neurological exam, especially MRI to diagnose this disease. MR images are the most valuable and main tool for identification multiple sclerosis and it is widely used for diagnosis and monitoring the MS patients [24]. There is no specific cure for MS, but medicines may slow it down and help control the disease progress [7]. Physical and occupational therapy may also be helpful. Because of the sensitivity of MR images, determine the disease clinically is hard and take a long time hence having an automatic method for diagnosing this disease fast and accurate is critical for decreasing the disease progression.

In this paper, different image processing methods have been used, such as edge detection and segmentation. Furthermore, some famous pre-processing techniques for having the best result have been used, like Brain extraction tool [13-16] and binarization. In addition, a modification method with Fuzzy C Means for better lesion segmentation and using canny for better edge detection has been presented. In section 3 we show that canny edge detection has a better result than other famous edge detection methods.  In our method, first edge detection method for finding edges has been used and then Fuzzy C-means has been applied for increasing diagnostic accuracy. In addition, for increasing the detection accuracy and having the best result, the clustering parameters have been changed and tested; also, as it was mentioned before, in MS lesion detection, effectiveness is the most important part. In this research, lesion diagnosis accuracy has been improved in comparison with others works [25].

International Journal of Artificial Intelligence and Applications (IJAIA), Vol.9, No.2, March 2018

## Authors

**Saba Heidari Gheshlaghi** was born in Tehran, Iran, in 1991. she received the B.Sc. degree in electrical engineering from the Shahid Rajaie University of Tehran in 2014. She is currently studying M.Sc. in control engineering from the Amirkabir University of Technology (Tehran Polytechnic). The areas of her research interests are image process, neuroscience, Cognitive science, and Neural Network.

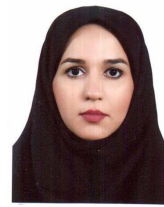

**Abolfazl Madani** was born in Arak, Iran, in 1992. He received the B.Sc. degree in electrical engineering from the Islamic Azad University of Arak in 2015. He is currently studying M.Sc. in control engineering from Islamic Azad University South Tehran Branch. The areas of his research interests are Neuroscience and Optimization methods for Deep Learning.

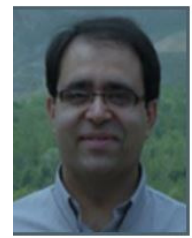

**Amir Abolfazl Suratgar** was born in Arak, Iran, in 1974. He received his B.S. degree in electrical engineering from Isfahan University of Technology, Isfahan, Iran, in 1996. He was honored M.S. and Ph.D. degrees in control engineering from Amirkabir University of Technology (Tehran Polytechnic), Tehran, Iran, in 1999 and 2002, respectively. He was an Assistant Professor of Electrical Engineering at the University of Arak, Arak, Iran. He received the outstanding research award from Engineering Faculty of University of Arak in 2006 and 2005. He received the outstanding education award from Engineering Faculty of University of Arak in 2009. Currently, he is an Associate Professor at

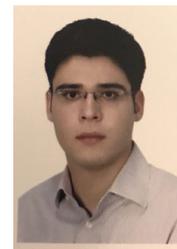

Amirkabir University of Technology (Tehran Polytechnic), Tehran, Iran. The areas of his research interests are computational intelligence, fuzzy modeling, fuzzy system stability analysis, neuroscience, and micro robots and MEMS dynamics analysis and control. He has published more than 100 papers in scientific journals and conferences.

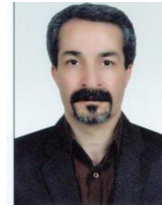

**Fardin Faraji** was born in Arak, Iran, in 1965. He received his Ph.D. degree in Neurology from Tehran University of Medical Sciences, Tehran, Iran. He is an Assistant Professor of Neurology at the University of Arak, Arak, Iran. He is the founder of MS association in Arak, Iran. The areas of his research interests are neurology and Multiple Sclerosis disease. He has published more than 45 papers in scientific journals and conferences.